\documentclass{acm_proc_article-sp}
\usepackage{etoolbox}
\usepackage{graphicx}
\usepackage{hanging}
\usepackage{hyperref}
\usepackage{mdframed}
\usepackage[frozencache]{minted}
\usepackage{subcaption}
\usepackage{tabularx}
\usepackage[labelfont=bf]{caption}
\makeatletter
\newcommand\subparagraph{%
  \@startsection{subparagraph}{5}
  {\parindent}
  {3.25ex \@plus 1ex \@minus .2ex}
  {-1em}
  {\normalfont\normalsize\bfseries}}
\makeatother
\usepackage{titlesec}
\let\subparagraph\relax 

\begin{document}

\titlespacing\section{0pt}{8pt plus 4pt minus 2pt}{5pt plus 1pt minus 1pt}
\titlespacing\subsection{0pt}{5pt plus 4pt minus 2pt}{5pt plus 1pt minus 1pt}

\BeforeBeginEnvironment{minted}{
  \begin{mdframed}[skipabove=15pt,
                   skipbelow=10pt,
                   linewidth=0pt,
                   innerleftmargin=5pt,
                   innertopmargin=0pt,
                   innerbottommargin=0pt]}
\AfterEndEnvironment{minted}{\end{mdframed}}

\conferenceinfo{}{Bloomberg Data for Good Exchange 2016, NY, USA}

\title{
  Themis-ml:
}
\subtitle{
    A Fairness-aware Machine Learning Interface for \\
    End-to-end Discrimination Discovery and Mitigation
}

\numberofauthors{1}
\author{
\alignauthor
Niels Bantilan\\
       \affaddr{Arena.io}\\
       \affaddr{New York, NY}\\
       \email{niels.bantilan@gmail.com}
}

\maketitle

\begin{abstract}
As more industries integrate machine learning into socially sensitive decision
processes like hiring, loan-approval, and parole-granting, we are at risk of
perpetuating historical and contemporary socioeconomic disparities. This is a
critical problem because on the one hand, organizations who use but do not
understand the discriminatory potential of such systems will facilitate the
widening of social disparities under the assumption that algorithms are
categorically objective. On the other hand, the responsible use of machine
learning can help us measure, understand, and mitigate the implicit historical
biases in socially sensitive data by expressing implicit decision-making mental
models in terms of explicit statistical models. In this paper we specify,
implement, and evaluate a ``fairness-aware'' machine learning interface called
themis-ml, which is intended for use by individual data scientists and
engineers, academic research teams, or larger product teams who use machine
learning in production systems.
\end{abstract}





\section{Introduction} In recent years, the transformative potential of machine
learning (ML) in many industries has propelled ML into the forefront of
mainstream media. From improving products and services to optimizing logistics
and operations, ML and artificial intelligence more broadly offer a wide range
of tools for organizations to enhance their internal and external capabilities.

As with any tool, we can use ML to engender great social benefit, but as
\cite{o2017weapons} emphasizes, we can also misuse it to bring about devastating
harm. In this paper, we focus on ML systems in the context of Decision Support
Systems (DSS), which are software systems that are intended to assist humans in
various decision-making contexts \cite{yoshimura2006decision,
montgomery2000evaluation, barnett1987dxplain, mysiak2005towards}. The misuse of
ML in these types of systems could potentially precipitate a widespread adverse
impact on society by introducing insidious feedback loops between biased
historical data and current decision-making \cite{o2017weapons}.

Researchers have developed many discrimination discovery and fairness-aware ML
methods \cite{kusner2017counterfactual, kamiran2012data, kamishima2012fairness,
kamiran2012decision, zemel2013learning, zafar2017fairness, dwork2012fairness,
zliobaite2015survey}, so we build on work done by others and seek to leverage
these techniques in the context of research- and product-based machine learning
applications.

Our contributions in this paper are three-fold. First, we propose an application
programming interface (API) for ``Fairness-aware Machine Learning Interfaces''
(FMLI) in the context of a simple binary classifier. Second, we introduce
themis-ml, an FMLI-compliant library, and apply it to a hypothetical
loan-granting DSS using the German Credit Dataset \cite{bache2013uci}. Finally, we
evaluate the efficacy of themis-ml as a tool for measuring potential
discrimination (PD) in both training data and ML predictions as well as
mitigating PD using fairness-aware methods. Our hope is that themis-ml serves as
a reference implementation that others might use and extend for their own
purposes.

\section{Bias and Discrimination}
Colloquially, bias is simply a preference for or against something, e.g.
preferring vanilla over chocolate ice cream. While this definition is
intuitive, here we explicitly define algorithmic bias as a form of bias that
occurs when mathematical rules favor one set of attributes over others in
relation to some target variable, like ``approving'' or ``denying'' a loan.

Algorithmic bias in machine learning models can occur when a trained model
systematically generates predictions that favor one group over another in
relation to some set of attributes, e.g. education, and some target variable,
e.g. ``default on credit''. While the definition above of bias is amoral,
discrimination is in essence moral, occurring when an action is based on biases
resulting in the unfair treatment of people. We define fairness as the inverse
of discrimination, meaning that a ``fairness-aware'' model is one that produces
non-discriminatory predictions.

Bias can lead to either direct (intended/explicit) or indirect
(unintended/implicit) discrimination, and the predominant legal concepts used to
determine these two types are known as disparate treatment and disparate impact,
respectively \cite{barocas2016big}. As \cite{kusner2017counterfactual,
kamiran2012data} suggest, we can address disparate treatment in ML models by
simply removing all variables that are highly correlated to the protected class
of interest, in addition to the protected class itself, from the training data.
However, as \cite{kusner2017counterfactual} points out, doing so does not
necessarily mitigate discriminatory predictions and may actually introduce
unfairness into an otherwise fair system. In contrast, addressing disparate
impact is more complex because it depends on historical processes that generated
the training data, non-linear relationships between the features and protected
class, and whether we are interested in measuring individual- or group-level
discrimination \cite{dwork2012fairness}.

\section{A Fairness-aware Machine Learning Interface}

So how does one measure
disparate impact and individual-/group-level discrimination in an ML-driven
product? In this section, we describe the main components of a simple
classification system, enumerate a few of the use cases that a research or
product team might have for using an FMLI, and propose an API that fulfills
these use cases.

A simple classification ML pipeline consists of five steps: data ingestion, data
preprocessing, model training, model evaluation, and prediction generation on
new examples. Data ingestion is outside the scope of this paper because it is a
highly variable process that depends on the application, often involves
considerable engineering effort, and potentially requires external stakeholder
buy-in.

Table 1 outlines a simple classification system in terms of the core interfaces
in scikit-learn (sklearn), which is a machine learning library in the Python
programming language \cite{buitinck2013api}, and table 2 delineates some of the
use cases that research or product teams might have to justify the use of an
FMLI.

\begin{table}
  \caption{A Simple Classification Pipeline}
  \renewcommand{\arraystretch}{1.75}
  \small\noindent\begin{tabularx}{\linewidth}{l X X}
    \textbf{API Interface} & \textbf{Function} & \textbf{Examples} \\
    \hline
    \textbf{Transformer} &
      \emph{Preprocess} raw data for model training. &
      mean-unit variance scaling, min-max scaling \\
    \textbf{Estimator} &
      \emph{Train} models to perform a classification task. &
      logistic regression, random forest \\
    \textbf{Scorer} &
      \emph{Evaluate} performance of different models. &
      accuracy, f1-score, area under the curve \\
    \textbf{Predictor} &
      \emph{Predict} outcomes for new data. &
      single-classifier prediction, ensemble prediction
  \end{tabularx}
\end{table}

\section{FMLI Specification}

Here we propose a high-level specification of themis-ml, an open source FMLI
named after the ancient Greek titaness of justice (the library can be found on
\href{https://github.com/cosmicBboy/themis-ml}{\underline{github}}.)  We adopt
sklearn's principles of consistency, inspection, non- proliferation of classes,
composition, and sensible defaults \cite{buitinck2013api}, and extend them with
the following FMLI-specific principles:

\begin{hangparas}{.15in}{1}
  \textbf{Model flexibility.} Focus on fairness-aware methods that are applicable
    to a variety of model types because users might have no control or full
    control over the specific model training implementation.
\end{hangparas}
\begin{hangparas}{.15in}{1}
  \textbf{Fairness as performance.} Provide estimators and scoring metrics that
    explicitly encode a notion of both model accuracy and fairness so that models
    can optimize for both.
\end{hangparas}
\begin{hangparas}{.15in}{1}
\textbf{Transparency of fairness-utility tradeoff.} Fair models often make less
  accurate predictions [8, 13], which is an important factor when assessing
  their business impact.
\end{hangparas}

\begin{table}
  \caption{FMLI Use Cases}
  \renewcommand{\arraystretch}{1.75}
  \small\noindent\begin{tabularx}{\linewidth}{X X}
    \textbf{Use Case} & \textbf{Rationale} \\
    \hline
    Detect and reduce discrimination in a production machine learning pipeline.
      & Fairness-aware modeling aligns with team/company values,
        provides protection from legal liability. \\
    Measure individual-/group-level discrimination in data with respect to a
    protected class and outcome of interest.
      & Need to assess the potential bias resulting from training models on
        data. \\
    Preprocess raw data or post-process model predictions in a way that reduces
    discriminatory predictions generated by models.
      & Unable to change the underlying implementation of the model training
        process. \\
    Explicitly learn model parameters that produce fair predictions for a variety
    of model types.
      & Need for flexibility when experimenting with or deploying different model
        types. \\
    Evaluate the degree to which fairness-aware methods reduce discrimination
    and assess the fairness-utility tradeoff.
      & Need for assessing the business consequences or other implications of
        deploying a fairness-aware model.
  \end{tabularx}
\end{table}

\subsection{Preliminaries} In the following subsections we describe specific
methods from the ML fairness literature that map onto each of the sklearn
interfaces. Note that we only provide a high level summary of each method,
citing the original sources for more implementation details. The following
descriptions make two assumptions: (i) the positive target label \(y^{+}\)
refers to a desirable outcome, e.g. ``approve loan'', and vice versa for the
negative target label \(y^{-}\), and (ii) the protected class is a binary
variable defined as \(s \in \{d, a\}\), where \(X_d\)  are members of the
disadvantaged group and \(X_a\) are members of the advantaged group.

Following these conventions, we define \(X_{d, y^{+}}\) and \(X_{d, y^{-}}\)
as the set of observations of the disadvantaged group that are positively
labelled and negatively labelled, respectively. Similarly, \(X_{a, y^{+}}\), and
\(X_{a, y^{-}}\)  are observations of the advantaged group that are positively
and negatively labelled, respectively.

\subsection{Transformer} The main idea behind fairness-aware preprocessing is to
take a dataset \(D\) consisting of a feature set \(X_{train}\), target labels
\(y_{train}\), and protected class \(s_{train}\) to output a modified dataset.

\emph{Relabelling}, also called \emph{Massaging}, modifies \(y_{train}\) by relabelling
the target variables in such a way that ``promotes'' members of the
disadvantaged protected class (e.g. ``immigrant'') and ``demotes'' members of
the advantaged class (e.g. ``citizen'') \cite{kamiran2012data}. A ranker \(R\)
(e.g. logistic regression) is trained on \(D\), and ranks are generated for
all observations. Some of the top-ranked observations \(X_{d, y^{-}}\)  are
``promoted'' to \(X_{d, y^{+}}\) and some of the bottom-ranked observations
\(X_{a, y^{+}}\)  are ``demoted'' to \(X_{a, y^{-}}\) such that the proportion
of \(y^{+}\) are equal in both \(X_d\) and \(X_a\). Two caveats of this method
are that it is intrusive because it directly manipulates \(y\), and that it
narrowly defines fairness as the uniform distribution of benefits between
\(X_a\) and \(X_d\).

\begin{minted}[mathescape,
               gobble=2,
               framesep=5mm,
               fontsize=\small]{python}
  from themis_ml.preprocess import Relabeller
  from sklearn.linear_model import LogisticRegression

  # use logistic regression as the ranking algorithm
  massager = Relabeller(ranker=LogisticRegression)

  # obtain a new set of labels
  new_y = massager.fit_transform(X, y, s)

  # train any model on new y labels
  lr = LogisticRegression()
  lr.fit(X, new_y)
\end{minted}

\emph{Reweighting} takes a dataset D and assigns a weight to each observation
using conditional probabilities based on \(y\) and \(s\) \cite{kamiran2012data}.
In brief, large weights are assigned to \(X_{d, y^{+}}\) and \(X_{a, y^{-}}\) ,
while small weights are assigned to \(X_{d, y^{-}}\) and \(X_{a, y^{+}}\). The
weights are then used as input to model types that support weighted sample
observations --- which actually points to the main limitation of this method,
since not all classifiers can incorporate observation weights during the
learning process.

\begin{minted}[mathescape,
               gobble=2,
               framesep=5mm,
               fontsize=\small]{python}
  from themis_ml.preprocess import Reweight
  from sklearn.linear_model import LogisticRegression

  reweigher = Reweight()

  # obtain fairness-aware weights for each observation
  reweigher.fit(y, s)
  fair_weights = reweigher.transform(y, s)

  # train a LogisticRegression model with sample weights
  lr = LogisticRegression()
  lr.fit(X, y, weights=fair_weights)
\end{minted}

\emph{Sampling} is composed of two methods: the first involves uniformly
sampling n observations from each group, where n is the expected size of that
group assuming a uniform distribution. The second is to preferentially sample
observations using a ranker \(R\), similar to the \emph{Relabelling} method. The
procedure is to duplicate the top-ranked \(X_{d, y^{+}}\) and \(X_{a, y^{-}}\)
while removing top-ranked \(X_{d, y^{-}}\) and \(X_{a, y^{+}}\)
\cite{kamiran2012data}.

\begin{minted}[mathescape,
               gobble=2,
               framesep=5mm,
               fontsize=\small]{python}
  from themis_ml.preprocess import (
      UniformSample, PreferentialSample)
  from sklearn.linear_model import LogisticRegression

  # use logistic regression as the ranking algorithm
  uniform_sampler = UniformSample()
  preferential_sampler = PreferentialSample(
      ranker=LogisticRegression)

  # obtain a new dataset with uniform sampling
  uniform_sampler.fit(y_train, s_train)
  X, y, s = uniform_sampler.transform(X, y, s)

  # obtain a new dataset with preferential sampling
  preferential_sampler.fit(y_train, s_train)
  X, y, s = preferential_sampler.transform(X, y, s)
\end{minted}

\subsection{Estimator}

Themis-ml implements two methods for training fairness-aware models: the
prejudice remover regularizer (PRR), and the additive counterfactually fair
(ACF) model.

\cite{kamishima2012fairness} proposes PRR as an optimization technique that
extends the standard L1/L2-norm regularization method \cite{ng2004feature,
ribeiro2016should} by adding a prejudice index term to the objective function.
This term is equivalent to normalized mutual information, which measures the
degree to which predictions \(y\) and \(s\) are dependent on each other. With
values ranging from 0 to 1, 0 means that \(y\) and \(s\) are independent, and a
value of 1 means that they are dependent. The goal of the objective function is
to find model parameters that minimize the difference between the true label
\(y\) and the predicted label \(\hat{y}\) in addition to the degree to which
\(y\) depends on s.

\begin{minted}[mathescape,
               gobble=2,
               framesep=5mm,
               fontsize=\small]{python}
  from themis_ml.linear_model import LogisticRegressionPRR

  # use L2-norm regularization and prejudice index as
  # the discrimination penalizer
  lr_prr = LogisticRegressionPRR(
      penalty="L2", discrimination_penalty="PI")

  # fit the models
  lr_prr.fit(X, y, s)
\end{minted}

ACF is a method described by \cite{kusner2017counterfactual} within the
framework of counterfactual fairness. The main idea is to train linear models to
predict each feature using the protected class attribute(s) as input. We can
then compute the residuals \(\epsilon_{ij}\) between the predicted feature
values and true feature values for each observation i and each feature j. The
final model is then trained on \(\epsilon_{ij}\) as features to predict y.

\begin{minted}[mathescape,
               gobble=2,
               framesep=5mm,
               fontsize=\small]{python}
  from themis_ml.linear_model import LinearACFClassifier

  # by default, LinearACFClassifier uses linear
  # regression as the continuous feature estimator
  # and logistic regression as the binary feature
  # estimator and target variable classifier
  linear_acf = LinearACFClassifier()

  # fit the models
  linear_acf.fit(X_train, y_train, s_train)
\end{minted}

\subsection{Predictor}

Themis-ml draws on two methods to make model type-agnostic predictions:
\emph{Reject Option Classification} (ROC) and \emph{Discrimination Aware
Ensemble Classification} (DAEC) \cite{kamiran2012decision}. Unlike the
Transformer and Estimator methods outlined above, ROC and DAEC do not modify the
training data or the training process. Rather, they postprocess predictions in a
way that reduces potentially discriminatory (PD) predictions.

\cite{kamiran2012decision} describes two ways of implementing ROC, starting with
ROC in a single classifier setting. ROC works by training an initial classifier
on \(D\), generating predicted probabilities on the test set, and then computing the
proximity of each prediction to the decision boundary learned by the classifier.
Within this boundary defined by the critical region threshold \(\theta\), where
0.5 < \(\theta\) < 1, \(X_d\) are assigned as \(y^{+}\) and \(X_a\) are assigned
as \(y^{-}\). ROC in the multiple classifier setting is similar to the single
classifier setting, except that predicted probabilities are defined as the
weighted average of probabilities generated by each classifier.

\begin{minted}[mathescape,
               gobble=2,
               framesep=5mm,
               fontsize=\small]{python}
  from themis_ml.postprocessing import (
      SingleROClassifier, MultiROClassifier)
  from sklearn.linear_model import LogisticRegression
  from sklearn.tree import DecisionTreeClassifier

  # use logistic regression for single classifier setting
  single_roc = SingleROClassifier(
      estimator=LogisticRegression())

  # use logistic regression and decision trees for
  # multiple classifier setting
  multi_roc = MultiROClassifier(
      estimators=[LogisticRegression(),
                  DecisionTreeClassifier()])

  # fit the models and generate predictions
  single_roc.fit(X, y, s)
  multi_roc.fit(X, y, s)
  single_roc.predict(X, s)
  multi_roc.predict(X, s)
\end{minted}

The main limitation of ROC is that model types must be able to produce predicted
probabilities. DAEC gets around this problem by training an ensemble of
classifiers and, through a similar relabelling rule as ROC, re-assigns any
prediction where classifiers disagree on the predicted label. As
\cite{kamiran2012decision} notes, in general, the larger the disagreement
between classifiers, the larger the reduction in discrimination.

\begin{minted}[mathescape,
               gobble=2,
               framesep=5mm,
               fontsize=\small]{python}
  from themis_ml.postprocessing import DAEnsembleClassifier
  from sklearn.linear_model import LogisticRegression
  from sklearn.tree import DecisionTreeClassifier

  # use logistic regression and decision trees
  dae_clf = DAEnsembleClassifier(
      estimators=[LogisticRegression(),
                  DecisionTreeClassifier()])

  # fit the models and generate predictions
  dae_clf.fit(X, y, s)
  dae_clf.predict(X, s)
\end{minted}

\subsection{Scorer}

The Scorer interface is concerned with measuring the degree to which data or
predictions are PD. Themis-ml implements two methods for measuring group-level
discrimination and two methods for measuring individual-level discrimination.

In the context of measuring group-level discrimination,
\cite{zliobaite2015survey} describes \emph{mean difference} and \emph{normalized
mean difference}. \emph{Mean difference} measures the difference between \(p(a
\cup y^{+})\) and \(p(d \cup y^{+})\). Values range from -1 to 1, where -1 is
the reverse-discrimination case (all \(X_a\) have \(y^{-}\) labels and all
\(X_d\) have \(y^{+}\) labels) and 1 is the fully discriminatory case
(all \(X_a\) have \(y^{+}\) labels and all \(X_d\) have \(y^{-}\) labels).
\emph{Normalized mean difference}, which also takes on values between -1 and 1,
scales these values based on the maximum possible discrimination in a dataset
given the rate of positive labels \cite{zliobaite2015survey}.

\begin{minted}[mathescape,
               gobble=2,
               framesep=5mm,
               fontsize=\small]{python}
  from themis_ml.metrics import (
      mean_difference, normalized_mean_difference)

  # compare group-level discrimination in true
  # labels and predicted labels
  md_y_true = mean_difference(y, s)
  md_y_pred = mean_difference(pred, s)
  md_y_pred - md_y_true

  norm_md_y_true = norm_mean_difference(y, s)
  norm_md_y_pred = norm_mean_difference(pred, s)
  norm_md_y_pred - norm_md_y_true
\end{minted}

\cite{zliobaite2015survey} also describes \emph{consistency} and \emph{situation
test score} as individual-level discrimination measures. \emph{Consistency}
measures the difference between the target label of a particular observation and
target labels of its neighbors. K-nearest neighbors (knn) measures the pairwise
distance between observations X. Then, for each observation \(x_i\) and each
neighbor \((x_j, y_j) \in knn(x_i)\), we compute the differences between \(y_i\)
and target labels of neighbor \(y_j\). A consistency score of 0 indicates that
there is no individual-level discrimination, and a score of 1 indicates that
there is maximum discrimination in the dataset.

The \emph{situation test score} metric is similar to \emph{consistency}, except
we consider only \(x_i \in X_d\). This method uses \emph{mean difference} to
compute a discrimination score among neighbors \(x_j \in knn(x_i)\), producing a
score between 0 and 1, where 0 indicates no discrimination, and 1 indicates
maximum discrimination \cite{zliobaite2015survey}.

\begin{minted}[mathescape,
               gobble=2,
               framesep=5mm,
               fontsize=\small]{python}
  from themis_ml.metrics import (
      consistency, situation_test_score)

  # compare individual-level discrimination
  # in true labels and predicted labels
  c_true = consistency(y, s)
  c_pred = consistency(y, s)
  c_pred - c_true

  sts_true = situation_test_score(y, s)
  sts_pred = situation_test_score(y, s)
  sts_pred - sts_true
\end{minted}

\section{Evaluating Themis-ml}

In this section we use the German Credit dataset \cite{bache2013uci} to evalute
themis-ml. We use \emph{mean difference} as the ``fairness'' measure and the
\emph{area under the curve} (AUC) as the ``utility'' measure. The former
represents the degree to which PD patterns in \(D\) are learned by the ML model,
and the latter represents the predictive power of a model given the available
dataset \((X, y, s) \in D\). The following analysis is by no means meant to be a
comprehensive investigation of all possible workflows that themis- ml enables.
However, does demonstrate the potential of themis-ml as a tool that facilites
fairness-aware machine learning by enabling the user to:

\begin{enumerate}
  \item Measure PD target label distributions in the training data.
  \item Measure PD predicted labels in a machine learning algorithm's predictions.
  \item Reduce PD predictions using fairness-aware techniques.
  \item Diagnose the fairness-utility tradeoff in a particular data context.
\end{enumerate}

The German Credit dataset classifies 1000 anonymized individuals as having
``good'' and ``bad'' credit risks as part of a bank loan application, which we
encode as \(1\) and \(0\) respectively to define the \emph{credit risk}
target variable.

Each individual is associated with twenty attributes such as the purpose of the
loan, employment status, and other personal information. We begin the analysis
by extracting three protected class attributes --- \emph{female},
\emph{foreign worker}, and \emph{age below 25} --- and encode them as binary
variables such that the putatively disadvantaged group is encoded as \(1\), and
the advantaged group is encoded as \(0\) (the advantaged group would be
\emph{male}, \emph{citizen worker}, and \emph{age above 25}, respectively).

Using the \textbf{Scorer} interface, we measure PD patterns with respect to
\emph{credit risk} and each of the protected classes defined above using the
\emph{mean difference} and \emph{normalized mean difference} metrics.

\textbf{Table 3} reports the PD distribution of ``good'' and ``bad'' credit
risks with respect to the protected attributes \emph{female}, \emph{foreign
worker}, and \emph{age below 25}. The fact that both the \emph{mean difference}
(md) and \emph{normalized mean difference} (nmd) scores are greater than zero
suggests that the probability of being classified as having ``good'' risk is
higher in the advantaged group than that of the disadvantaged group.

\begin{table}
  \caption{Potentially discriminatory target variable distribution.
    \emph{md} = mean difference, \emph{nmd} = normalized mean difference.}
  \renewcommand{\arraystretch}{1.75}
  \small\noindent\begin{tabularx}{\linewidth}{l|X|X|X|X}
    \textbf{protected class} & \textbf{md (\%)} & \textbf{md 95\% CI} &
      \textbf{nmd (\%)} & \textbf{nmd 95\% CI}\\
    \hline
    \textbf{female} & 7.48 & (1.35, 13.61) & 7.73 & (1.39, 14.06) \\
    \textbf{foreign worker} & 19.93 & (4.91, 34.94) & 63.96 & (15.76, 112.17)\\
    \textbf{age below 25} & 14.94 & (7.76, 22.13) & 17.29 & (8.97, 25.61)\\
  \end{tabularx}
\end{table}

\subsection{Experimental Procedure}

To assess the extent to which (i) a model trained on these data mirrors these PD
\emph{credit risk} distributions, and (ii) fairness-aware techniques can reduce
these methods, we used \emph{mean difference} to measure model fairness and
\emph{AUC} to measure model utility. For this experiment we specify five
conditions:

\begin{itemize}
  \item Baseline (\emph{B}): Train a model on all available input variables in
        the German Credit dataset, including protected attributes.
  \item Remove Protected Attribute (\emph{RPA}): Train a model on input variables
        without protected attributes. This is the naive fairness-aware approach.
  \item Relabel Target Variable (\emph{RTV}): Train a model using the
        \emph{Relabelling} fairness-aware method.
  \item Counterfactually Fair Model (\emph{CFM}): Train a model using the
        \emph{Additive Counterfactually Fair} method.
  \item Reject-option Classification (\emph{ROC}): Train a model using the
        \emph{Reject-option Classification} method.
\end{itemize}

For each of these conditions, we train LogisticRegression, DecisionTree, and
RandomForest model types using 10-fold cross validation; generate train and test
predictions; and compute \emph{AUC} and \emph{mean difference} metrics for each
train-test pair. We then compute the mean of these metrics for each condition and
model type. The code for this analysis is available on
\href{https://github.com/cosmicBboy/themis-ml/blob/master/paper/Evaluating%20Themis-ml.ipynb}{\underline{github}}.

\subsection{Measuring and Mitigating Potentially Discriminatory Predictions}

\begin{figure}[h]
\centering
\includegraphics[width=8.5cm]{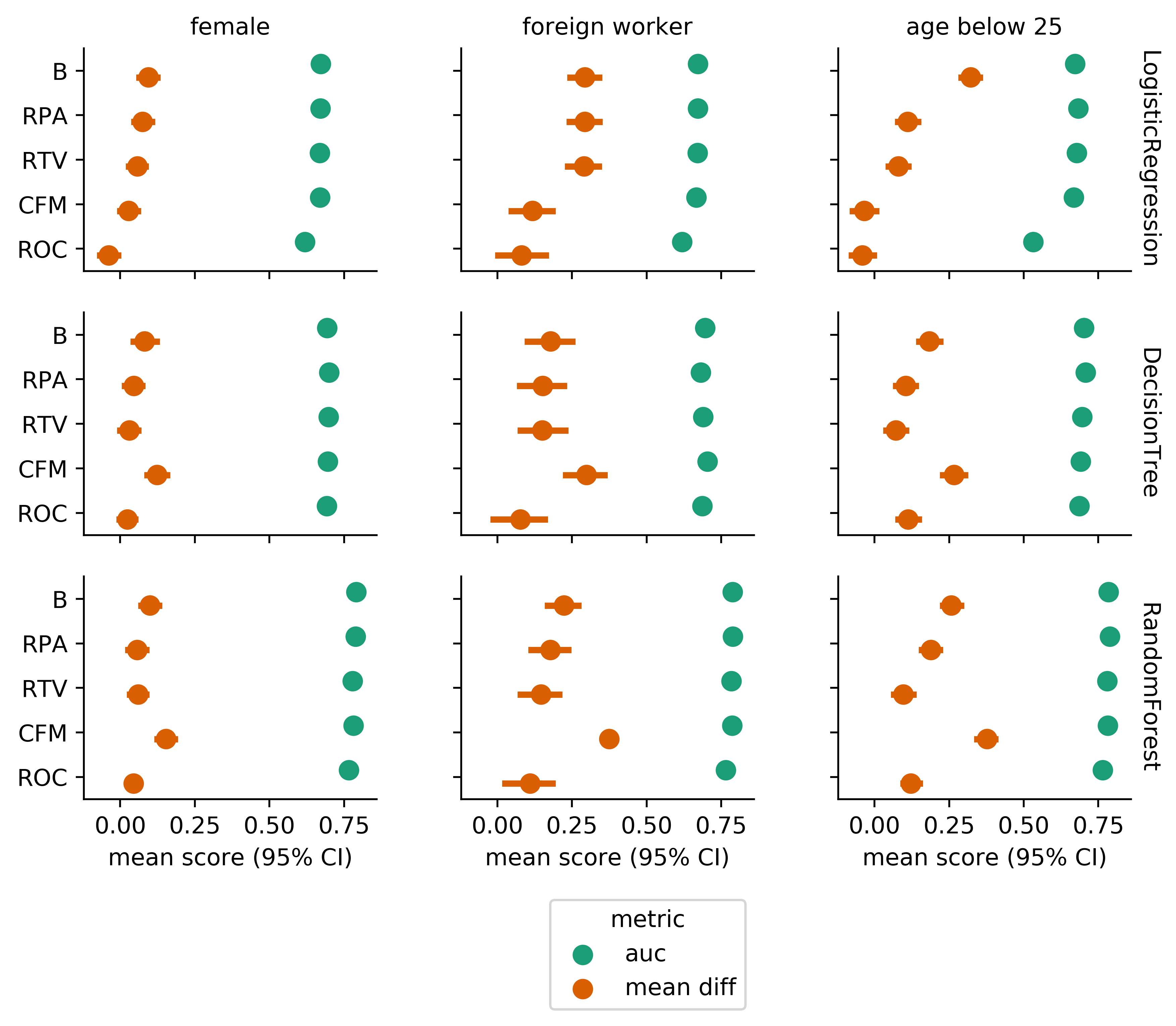}
\caption{
  \textbf{Comparison of Fairness-aware Methods} using LogisticRegression,
  DecisionTree, and RandomForest (rows) as base estimators for each protected
  attribute context (columns), measured by \emph{AUC} and
  \emph{mean difference} evaluated on test set predictions.}
\label{figure-1-comparison-fa-methods}
\end{figure}

Figure 1 suggests that in the case of LogisticRegression, the baseline model
\emph{B} does indeed mirror the PD patterns found in the true target variable.
Furthermore, each of the fairness-aware methods appear to have the desired
effect of reducing \emph{mean difference}, but to varying degrees depending on
the method and protected attribute. In the \emph{female} protected attribute
context, where there appears to be the least PD (mean difference of \(7.48\%\)),
the reductive effect of the fairness-aware methods do not appear to be as
large as in the \emph{foreign worker} and \emph{age below 25} contexts.

The lack of reduction in \emph{mean difference} between \emph{B} and \emph{RPA},
with respect to \emph{foreign worker} and LogisticRegression, illustrates the
observation made by \cite{kusner2017counterfactual} that removing protected
attributes from the training data does not necessarily prevent the algorithm
from mirroring PD patterns in the data.

However, the sizeable reduction in \emph{mean difference} between \emph{B} and
\emph{RPA}, with respect to \emph{age below 25} and LogisticRegression model,
shows that removing protected attributes can sometimes make models more fair
while also retaining predictive power.

An interesting thing to note here is that the \emph{Additive Counterfactually
Fair} method actually increases \emph{mean difference} for DecisionTrees and
RandomForests across all protected attribute contexts. Two possible explanations
behind this observation is that certain assumptions made by \emph{ACF} are not
suitable for non-linear learning algorithms, or the meta-estimators that compute
the residuals for non-linear estimators should be non-linear as well. This
is an open question worth future inquiry.

\subsection{The Fairness-utility Tradeoff}

Just as the bias-variance tradeoff has become a useful diagnostic tool to guide
ML research and application \cite{fortmann2012understanding}, the fairness-
utility tradeoff can help machine learning practitioners and researchers
determine which fairness-aware methods are suitable for their particular data
context.

\begin{figure}[h]
\centering
\includegraphics[width=8.5cm]{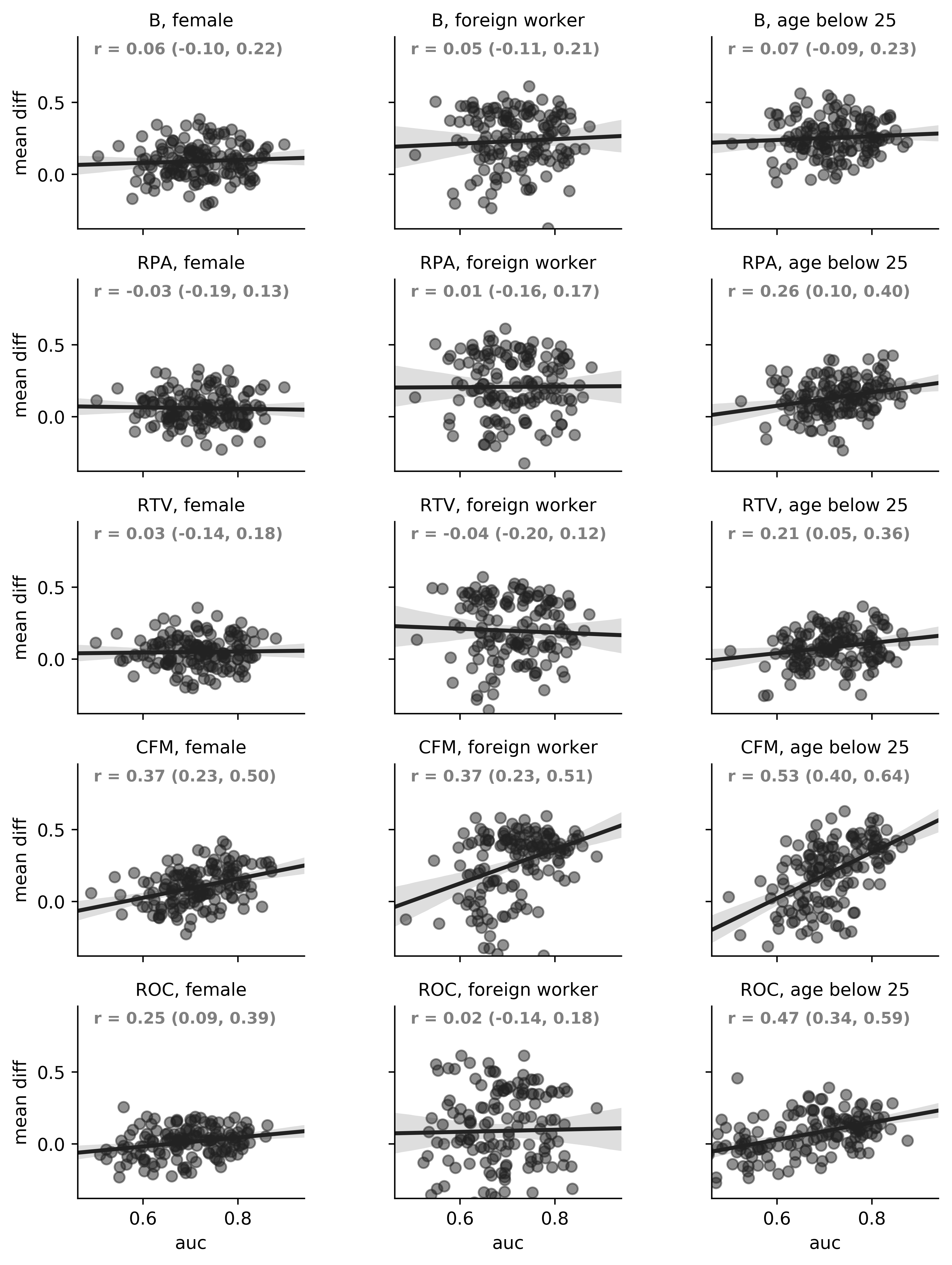}
\caption{
  \textbf{Correlation between AUC and Mean Difference} for each fairness-aware
  condition (rows) and protected attribute contexts (columns) across all model
  types (LogisticRegression, DecisionTree,   RandomForest). 95\% confidence
  intervals are provided for the pearson \(r\) correlation metric.}
\label{figure-2-fairness-utility-tradeoff}
\end{figure}

In figure 2, we visualize the fairness-utility tradeoff, in this case as
measured by \emph{mean difference} and \emph{AUC}, respectively. We report
pearson correlation coefficients \(r\) for each protected attribute context and
fairness-aware condition with their respective 95\% confidence intervals.

These results suggest that the relationship between fairness and utility is
noisy, however there does seem to be a consistent but weak positive correlation
between \emph{mean difference} and \emph{AUC} (or a negative correlation between
fairness and utility, since lower scores are better for \emph{mean difference}
and higher scores are better for \emph{AUC}).

Interestingly, we note the cases in which there are zero or negative \(r\)
coefficient values. \(r = 0\) implies that there is no tradeoff between fairness
and utility: one can expect to increase the utility of a set of models without
adversely affecting the fairness of predictions generated by those models.
Although there are no cases where \(r_{ci\ upper} < 0\), \(r < 0\) suggests that
it might be plausible to find regimes in which one can expect to increase both
the utility and fairness of a model. Future work in this area might examine the
asymptotic behavior of the relationship between fairness and utility as model
complexity increases.

Depending on one's use cases, analyses like this might prove to be a useful
guide for figuring out what kinds of methods are robust in the sense that one
can reduce PD predictions with little to no adverse
impact on predictive performance.

\section{Discussion}

In this paper, we describe and evaluate an FMLI in the classification context
where we consider only a single binary protected class variable and a binary
target variable.

More work needs to be done to generalize FMLIs to the multi-classification,
regression, and multiple protected classes settings. Furthermore, many basic
questions about model tuning, evaluation, and selection in the fairness-aware
context remain. For instance, what might be some reasonable ways to aggregate
utility and fairness metrics in order to find the optimal set of
hyperparameters? Additionally, little is understood about the composability of
fairness-aware methods, i.e., when different techniques are used together in
sequence, are the resulting discrimination reductions additive or otherwise?

Future technical work might also extend the FMLI specification to include
techniques like Locally Interpretable Model-Agnostic Explanations
\cite{ribeiro2016should} and develop legal frameworks for thinking about how
different stakeholders would interact with FMLIs. For example, companies that
choose not to expose the model-training components of their internal ML
pipeline could still grant some form of access to the predictions generated by
the models if there were to be a set of standards for model transparency and
accountability.

Finally, many of the fairness-aware methods, such as the \emph{Relabeller},
implicitly define fairness as the uniform (equal) distribution of benefits among
disadvantaged and advantaged groups. Future work would make this definition more
flexible, for example, by defining fairness as the proportional distribution of
benefits based on need. This would necessitate the mathematical formalization of
another set of assumptions about the needs of disadvantaged and advantaged
groups.

Given the challenges ahead, our ability to measure and mitigate discrimination
is limited by our common social, legal, and political understanding of fairness
itself. This common understanding is often lacking because marginalized social
groups typically do not have a voice at the table when defining what counts as
fair. Since FMLIs are simply a tool to measure and mitigate formalized
definitions of discrimination, it is important for all stakeholders to engage in
an inclusive forum where everyone, especially disadvantaged social groups, can
contribute.

\nocite{*}
\bibliographystyle{ieeetr}
\bibliography{references}

\end{document}